\documentclass[10pt]{article}
\usepackage{a4wide}

\def\beq{\begin{eqnarray}} 
\def\eeq{\end{eqnarray}}

\begin{document}

\parindent0pt

\begin{flushright}
{\small
WUB 98--40 \\
ADP-98-75/T342\\
HD-THEP-98-59  \\
hep-ph/9812269
}
\end{flushright}

\begin{center}
\vskip 2.0\baselineskip
\textbf{\Large 
Mixing and Decay Constants of Pseudoscalar Mesons:\\
The Sequel}
\vskip 2.5\baselineskip
Th.~Feldmann$^{1}$, P.~Kroll$^{1,2}$ and B.~Stech$^{3}$
\vskip \baselineskip
{\small
1. Fachbereich Physik, Universit\"at Wuppertal, 
 D-42097 Wuppertal, Germany \\[0.3em] 
2. Centre for the Subatomic Structure of Matter, 
   University of Adelaide, SA 5005, Australia \\[0.3em]  
3. Institut f\"ur Theoretische Physik, Universit\"at Heidelberg, 
D-69120 Heidelberg, Germany
}
\vskip 2.5\baselineskip
\textbf{Abstract} \\[0.5\baselineskip]
\parbox{\textwidth}{ We present further tests and applications of the
new $\eta$-$\eta'$ mixing scheme recently proposed by us. The particle
states are decomposed into orthonormal basis vectors in a light-cone
Fock representation. Because of flavor symmetry breaking the mixing
of the decay constants can be identical to the mixing of particle
states at most for a specific choice of this basis. Theoretical and
phenomenological considerations show that the quark flavor basis has
this property and allows, therefore, for a reduction of the number of
mixing parameters. A detailed comparison with other mixing schemes is
also presented. }

\vskip 1.5\baselineskip
\end{center}


In a recent reinvestigation \cite{FeKrSt98} of processes involving
$\eta$ and $\eta'$ mesons we pointed out that a proper treatment of
the $\eta$-$\eta'$ system requires a sharp distinction between the
mixing properties of the meson states from the mixing properties of
the decay constants. While the particle state mixing involves the
global wave functions, the decay constants probe the quark
distributions at zero spatial separation. Conventionally, $\eta$ and
$\eta'$ are expressed as superpositions of an $SU(3)$ flavor octet and
a flavor singlet state corresponding to an orthogonal transformation
with mixing angle $\theta$. The decay constants of $\eta$ and $\eta'$
defined by their matrix elements with singlet and octet axial vector
currents will in general not show the same mixing since flavor
symmetry breaking manifests itself differently at small and large
distances. Because $SU(3)$ breaking is solely caused by the current
quark masses a simpler picture can be expected for properly defined
decay constants in the quark-flavor basis. Indeed, a dramatic
simplification is achieved by taking two orthogonal basis states
\footnote {In principle, the two-state basis should be extended by
states of higher energy, for instance by adding a $c\bar c$ state.
Energy considerations indicate, however, that the mixing with these
states is small. The small charm components in the $\eta$ and $\eta'$
is discussed in \cite{FeKrSt98}.} which are assumed to have in a Fock
state description the parton composition 
\beq 
|\eta_q \rangle &=&
\Psi_q^{} \, |u\bar u + d\bar d\rangle/\sqrt2 + \ldots 
\cr 
|\eta_s \rangle &=& 
\Psi_s^{} \, |s\bar s \rangle + \ldots
\label{Eq1}
\eeq Here $\Psi_i$ denote (light-cone) wave functions of the
corresponding parton states, and the dots stand for higher Fock states
which also include $|gg\rangle$ components. These higher Fock states
play no explicit role in the following discussions where we are mainly
interested in mesonic states and decay constants. The physical meson
states are related to the basis (\ref{Eq1}) by an orthogonal
transformation
\beq 
\left( 
 \matrix{ 
 |\eta\phantom{{}'}\rangle \cr
 |\eta'\rangle
 }
\right) &=& 
U(\phi) \, \left(
 \matrix{ 
 |\eta_q\rangle \cr 
 |\eta_s\rangle
 }
\right) \ , 
\qquad U(\phi) = \left( 
 \matrix{
 \cos\phi & -\sin\phi \cr 
 \sin\phi & \phantom{-}\cos\phi 
 }
\right) \ ,
\label{Eq2}
\eeq where $\phi$ is the mixing angle. {\it Ideal mixing} corresponds
to the case $\phi=0$. It is to be emphasized that our definition of
meson states is given in terms of parton degrees of freedom without
introducing model-dependent concepts like constituent quarks.

Our central ansatz (\ref{Eq1},\ref{Eq2}) has important consequences
for the weak decay constants which probe the short-distance properties
of the quark-antiquark Fock states. To see this in detail, let us
define the decay constants\footnote{We stress that occasionally used
decay constants ``$f_\eta,f_{\eta'}$'' are ill-defined quantities.}
by ($f_\pi = 131$~MeV)
\beq 
&& \langle 0 | J_{\mu5}^i | P\rangle
 \equiv 
 i \, f_P^i \, p_\mu \ ,
\label{Eq3}
\eeq where $ P=\eta,\eta'$. Here $J_{\mu5}^i$ denotes the
axial-vector currents with quark content $i=q,s$. The decay constants
are related to the quark-antiquark wave functions at the origin of
configuration space. Because of the fact that light-cone wave
functions do not depend on the hadron momentum we can define two
basic decay constants $f_q$ and $f_s$ arising from $\eta_q$ and
$\eta_s$, respectively,
\beq 
f_i &=& 
2 \sqrt{6} \int \frac{dx \, d^2k_\perp}{16 \pi^3} \, 
 \Psi_i(x,k_\perp)
\label{origin} \ .
\eeq Here $x$ denotes the usual (light-cone~+) momentum fraction of
the quark and $k_\perp$ its transverse momentum with respect to its
parent meson's momentum. Eq.~(\ref{origin}) is exact, only the
quark-antiquark Fock state contributes to the decay constant, higher
Fock states do not contribute. Using Eqs.~(\ref{Eq1}--\ref{Eq3}), one
immediately observes that our ansatz for the Fock decomposition
naturally leads to decay constants in the quark-flavor basis which
simply follow the pattern of state mixing: 
\beq 
\left(
 \begin{array}{cc} 
 f_\eta^q & f_\eta^s \cr 
 f_{\eta'}^q & f_{\eta'}^s
 \end{array} 
\right) 
&=& 
U(\phi) \, {\rm diag}[f_q,f_s] \ .
\label{Eq4}
\eeq 


The conventional octet-singlet basis states are obtained from the
quark-flavor basis states by performing a rotation with the ideal
mixing angle. The physical states
are then related to the octet-singlet basis states by 
\beq 
\left(
 \matrix{ 
 |\eta\phantom{{}'}\rangle \cr 
 |\eta'\rangle
 }
\right) 
&=&
U(\theta) \, \left(
 \matrix{ 
 |\eta_8\rangle \cr 
 |\eta_1\rangle
 }
\right)
\label{eq3} \ ,
\eeq with $\theta=\phi-\arctan\sqrt2$. The corresponding Fock
decompositions of these octet-singlet basis states, following from
Eqs.~(\ref{Eq1},\ref{Eq2},\ref{eq3}), read 
\beq 
|\eta_8 \rangle &=&
 \frac{\Psi_q + 2 \, \Psi_s}{3} \, 
 \frac{|u\bar u + d\bar d - 2 s\bar s\rangle}{\sqrt6} 
+ \frac{\sqrt2 \, (\Psi_q - \Psi_s)}{3} \,
 \frac{|u\bar u + d\bar d + s\bar s\rangle }{\sqrt3} 
+ \ldots
\nonumber \\[0.2em] 
|\eta_1 \rangle &=& 
 \frac{\sqrt2 \, (\Psi_q - \Psi_s)}{3} \, 
 \frac{|u\bar u + d\bar d - 2 s\bar s\rangle}{\sqrt6} 
+ \frac{2 \,\Psi_q + \Psi_s}{3} \, 
 \frac{|u\bar u + d\bar d + s\bar s\rangle}{\sqrt3} 
+ \ldots
\label{81def}
\eeq Obviously, it is unavoidable that the so-defined octet (singlet)
meson state contains an $SU(3)$ singlet (octet) admixture, except
for identical wave functions $\Psi_q=\Psi_s$, an equality which holds
in the flavor symmetry limit only. Only then one would find pure
octet and singlet states in Eq.~(\ref{81def}). Certainly, these
results are based on our central ansatz, namely that $\eta$ and
$\eta'$ can be decomposed into two orthogonal states where one state
has no $s\bar s$ and the other no $q\bar q$ component. One may
alternatively start from the assumption that $\eta_8$ and $\eta_1$
defined in Eq.~(\ref{eq3}) have Fock decompositions with only parton
octet or singlet combinations in the quark-antiquark sector,
respectively. Rotating these states back by the ideal mixing angle,
the resulting $\eta_q$ ($\eta_s$) state has an $s\bar s$ ($q\bar q$)
component, unless the octet and singlet wave functions are equal.
However, from both, theoretical and phenomenological considerations
performed in \cite{FeKrSt98,Leutwyler97,FeKrFF}, the quark-flavor
basis is to be favored. 

One may also define decay constants through
matrix elements of octet and singlet axial-vector currents,
analogously to Eq.~(\ref{Eq3}). Using Eqs.~(\ref{Eq1},\ref{Eq2}), one
easily sees that these decay constants cannot be expressed as $
U(\theta) \, {\rm diag}[f_8,f_1]$. Rather one has 
\beq 
\left(
 \begin{array}{cc}
 f_\eta^8 & f_\eta^1 \cr 
 f_{\eta'}^8 & f_{\eta'}^1
 \end{array}
\right) 
&=& 
\left( 
 \begin{array}{cc} 
 f_8 \,\cos\theta_8 & -f_1 \, \sin\theta_1 \cr
 f_8 \,\sin\theta_8 & \phantom{-} f_1 \, \cos\theta_1 
 \end{array} 
\right) \,
\label{tas} \ ,
\eeq where we use the new and general parametrization introduced in
Ref.~\cite{Leutwyler97}. The parameters appearing in Eq.~(\ref{tas})
are related to the basic parameters $\phi$, $f_q$ and $f_s$,
characterizing the quark flavor mixing scheme as follows
\cite{FeKrSt98}, 
\beq 
\theta_8 &=& \phi - \arctan\frac{\sqrt2 f_s}{f_q}, \quad 
 f_8^2 = \frac{f_q^2+2f_s^2}{3} , 
\cr 
\theta_1 &=& \phi - \arctan\frac{\sqrt2 f_q}{f_s}, \quad 
 f_1^2 = \frac{2f_q^2+f_s^2}{3} \ .
\label{eqrel}
\eeq 
The decay constants $f_P^i$ do not follow the pattern of state
mixing in the octet-singlet basis; only in the $SU(3)_F$ symmetry
limit one would have $\theta_8 = \theta = \theta_1$. This is a
consequence of the non-trivial Fock decomposition in
Eq.~(\ref{81def}). The difference between $\theta_8$ and $\theta_1$
following from Eq.~(\ref{eqrel}) leads to the same formula as derived
within chiral perturbation theory \cite{Leutwyler97}. In our approach
the quantities $\theta_8$ and $\theta_1$ are parameters determined by
the fundamental quantities\footnote{ We remark at this point, that the
flavor singlet axial-vector current is not conserved in QCD.
Consequently, the singlet decay constant $f_1$, and, hence, $f_q$ and
$f_s$ too, are renormalization scale dependent, although only mildly
since the corresponding anomalous dimension is of order $\alpha_s^2$
\cite{Leutwyler97}. Varying the scale $\mu$ between $M_\eta$ and
$M_{\eta_c}$, the value of $f_1(\mu)$ changes by 3\% only, an effect
which we discard.} $\phi$, $f_q$ and $f_s$. They are {\em not} to be
used as $|\eta\rangle = \cos\theta_8 |\eta_8\rangle - \sin\theta_1
|\eta_1\rangle$ etc.


Let us now briefly review the determination of the mixing
parameters performed in Ref.~\cite{FeKrSt98}. 
For this purpose we considered the divergences
of axial-vector currents which incorporate the $U(1)_A$ anomaly
($q_i=u,d,s$)
\begin{equation}
\partial^\mu \, \bar q_i \, \gamma_\mu\gamma_5 \, q_i 
 = 2 m_i \, \bar q_i \, i \gamma_5 \, q_i 
+ \frac{\alpha_s}{4\pi} \, G \tilde G \ , 
\label{anomaly}
\end{equation}
Sandwiching Eq.~(\ref{anomaly})
between the vacuum and the meson states and using the
definition of the decay constants (\ref{Eq3})
together with (\ref{Eq4}) one obtains the
$\eta$-$\eta'$ mass matrix in the quark flavor basis. 
Its elements are composed
of the gluonic matrix elements
$\langle 0 | \frac{\alpha_s}{4\pi} \, G \tilde G |\eta_i\rangle$
and matrix elements of the quark mass terms contained in (\ref{anomaly}), 
which can be expressed by the pion
and the $K$ meson masses using flavor symmetry and its breaking to first
order. The known eigenvalues of this mass matrix (the masses of $\eta$ and
$\eta'$) give then the value of the gluonic matrix elements
and, in particular, the value $42.4^\circ$ for the mixing angle $\phi$.

Alternatively, the mixing parameters can be determined from
phenomenology without using $SU(3)_F$ relations. 
The mixing angle $\phi$ can be determined by considering
appropriate ratios of decay widths or cross sections, in which either
the $\eta_q$ or the $\eta_s$ components are probed. The decay
constants $f_q$ and $f_s$ can be evaluated from the
$\eta,\eta'\to\gamma\gamma$ decay widths, relying on the chiral
anomaly prediction. The analysis of a number of decay and scattering
processes leads to the phenomenological set of parameters 
$f_q/f_\pi=1.07$, $f_s/f_\pi=1.34$, $\phi=39.3^\circ$ which we
will use in the following.
We like to point out that the phenomenological
values for the mixing angle $\phi$ from different experiments are
all consistent with each other within a rather small uncertainty.
The resulting differences between
$\theta_8$, $\theta$ and $\theta_1$ (although only caused by $SU(3)_F$
breaking effects) are enormous, see Table~\ref{Comptab}. We also
list in this table (in anti-chronological order) the parameter values
obtained in previous approaches.

\begin{table*}[t]
\begin{center}
\begin{tabular}{ ccccc || l}
$\theta$ & $\theta_8$ & $\theta_1$ & $f_8/f_\pi$ & $f_1/f_\pi$ & method 
\\
\hline \hline
$-12.3^\circ$ & $-21.0^\circ$ & $-2.7^\circ$ & $1.28$ & $1.15$ & 
 $qs$--scheme (theo.\,) \cite{FeKrSt98}
\\ 
$-15.4^\circ$ & $-21.2^\circ$ & $-9.2^\circ$ & $1.26$ & $1.17$ & 
 $qs$--scheme (phen.) \cite{FeKrSt98}
\\
\hline
-- & $-20.5^\circ$ & $-4^\circ$ & $1.28$ & $1.25$ & 
 ChPT \cite{Leutwyler97}
\\ 
$-21.4^\circ$ & x & x & $1.19$ & $1.10$ & 
 GMO formula \cite{Burakovsky:1998vc}
\\ 
$-15.5^\circ$ & -- & -- & -- & -- & 
 phenomenology \cite{Bramon97} 
\\ 
$-19.7^\circ$ & $[-12.2^\circ]$ & $[-30.7^\circ]$ & $[0.71]$ & $[0.94]$ & 
 model \cite{KiPe93}
\\ 
$-12.6^\circ$ & $[-19.5^\circ]$ & $[-5.5^\circ]$ & $[1.27]$ & $[1.17]$ & 
 model \cite{Schechter:1993iz}
\\
$-(23^\circ-17^\circ)$ & x & x & $1.2-1.3$ & $1.0-1.2$ & 
 phenomenology \cite{GaLe85,DoHoLi86,GiKa87,BaFrTy95}
\\ 
$-9^\circ$ & $[-20^\circ]$ & $[-5^\circ]$ & $[1.2]$ & $[1.1]$ & 
 $U(1)_A$ anomaly \cite{Diakonov:1981nv} 
\end{tabular}
\end{center}
\caption{Comparison of different determinations of mixing parameters.
The values given in parentheses are not quoted in the original
literature but have been evaluated by us from information given
therein. Crosses indicate approaches where the difference between
$\theta$, $\theta_1$ and $\theta_8$ has been ignored.}
\label{Comptab}
\end{table*}


It is possible to take decompositions of $\eta$ and $\eta'$ where
{\em either} the octet {\em or} the singlet state is pure -- in the
sense that admixtures of the orthogonal quark-antiquark combination
are absent -- but not both. Note that
one may, for instance, write the weak decay constants in 
Eq.~(\ref{tas}) in the form
\beq 
\left( 
 \begin{array}{cc}
 f_\eta^8 & f_\eta^1 \cr 
 f_{\eta'}^8 & f_{\eta'}^1
 \end{array}
\right) 
&=& U(\theta_8) \, \left( 
 \begin{array}{cc} 
 f_8 & f_1 \, \sin(\theta_8-\theta_1) \cr
 0 & f_1 \, \cos(\theta_8-\theta_1)
 \end{array} 
\right) \ .
\label{t8rel}
\eeq
Then it is tempting to introduce a 
new basis by
\beq
\left(
 \matrix{
 |\eta\phantom{{}'}\rangle \cr 
 |\eta'\rangle
 }
\right) 
&=&
U(\theta_8) \, \left(
 \matrix{
 |\tilde \eta_8\rangle \cr 
 |\tilde \eta_1\rangle
 } 
\right) \ .
\label{tilde}
\eeq 
The elements of the second matrix on
the r.h.s.\ of Eq.~(\ref{t8rel}) can now be viewed as 
the decay constants of $\tilde \eta_8$, $\tilde \eta_1$
through octet or singlet axial vector currents, respectively.
This matrix is still non-diagonal but triangular.
The new basis 
has the special feature that the anomaly contributes to the 
singlet ($\tilde\eta_1$) mass alone, i.e. one has
\beq
\langle 0 | \frac{\alpha_s}{4\pi}G\tilde G|\tilde \eta_8\rangle&=&0 \ .
\label{singldom}
\eeq
This property is related to the fact that the ratio 
$\langle 0 | \frac{\alpha_s}{4\pi} \, G \tilde G | \eta'\rangle/
 \langle 0 | \frac{\alpha_s}{4\pi} \, G \tilde G | \eta\rangle$,  is given
by $-\cot\theta_8$ \cite{FeKrSt98}.
It allows to determine a Gell-Mann--Okubo formula for
the mass of the $\tilde\eta_8$ basis state.
Transforming the mass matrix found in the quark flavor basis (see above)
to the new basis (\ref{tilde}) one finds 
\beq
f_8^2 \, m_{\tilde 8\tilde 8}^2 &=&
 \frac{f_q^2 \, M_\pi^2+2 f_s^2 \, (2\, M_K^2-M_\pi^2)}{3} 
\label{gmotilde} \ .
\eeq 
This formula reminds of the suggestion put forward in
Ref.~\cite{Burakovsky:1998vc} (see also \cite{Donoghue}), namely to
use the product $f^2 M^2$ rather than $M^2$ to determine the $SU(3)_F$
breaking effects in the Gell-Mann--Okubo formula\footnote{Note however
that the decay constants used in the analysis of
\cite{Burakovsky:1998vc} are not defined as proper matrix elements of
weak currents.}. 
Insertion of the relations in
Eq.~(\ref{eqrel}) into Eq.~(\ref{gmotilde}) yields
\beq
m_{\tilde 8 \tilde 8}^2 
&\simeq& 
\frac{4 M_K^2 - M_\pi^2}{3} - 
 \Delta_{\rm GMO} \, \frac{M_K^2-M_\pi^2}{3}
\label{gmoexp}
\eeq with $\Delta_{\rm GMO} = 4 \, (f_q^2-f_s^2)/(3\,f_8^2)$. 
The deviation from the standard  Gell-Mann--Okubo
relation $\Delta_{\rm
GMO}$ can also be derived in chiral perturbation theory
\cite{LaPa74,GaLe85}.

The above discussion clearly shows: An analysis which implicitly uses
equation (\ref{singldom}) provides for an estimate of the parameter
$\theta_8$ rather than the angle $\theta$. Indeed, previous
treatments along these lines obtained mixing angles close to
$-20^\circ$, which is consistent with our value of $\theta_8$ (see
Table~\ref{Comptab}).

As a first test of our mixing approach we analyzed the $\eta\gamma$
and $\eta'\gamma$ transition form factors in Ref.~\cite{FeKrSt98}. A
good description of the experimental data has been found from the
phenomenological set of parameters. For details we refer to
\cite{FeKrSt98,FeKrFF}. Let us now turn to further tests and
applications of our results which have not been discussed in
Ref.~\cite{FeKrSt98}.


\vskip1em\noindent {\it Radiative Decays of $S$-wave quarkonia:}

We define the ratio of decay widths $R({}^3S_n) = \Gamma[{}^3S_n \to
\eta' \gamma]/ \Gamma[{}^3S_n \to \eta\gamma]$ where ${}^3S_n$
represents one of the quarkonia $J/\psi,\psi',\Upsilon,\ldots$
According to \cite{Novikov:1980uy} the photon is emitted by the $c$
quarks which then annihilate into lighter quark pairs through the
effect of the anomaly. Thus, the creation of the corresponding light
mesons is controlled by the matrix element $\langle
0|\frac{\alpha_s}{4\pi} G\tilde{G}|P\rangle$, leading to
\cite{FeKrSt98}
\beq 
R({}^3S_n) &=& 
\cot^2 \theta_8 \, \left(\frac{k_{\eta'\gamma}}{k_{\eta\gamma}}\right)^3
\label{rv}
\eeq where $k_{12}=\sqrt{(M^2-m_1^2-m_2^2)^2-4m_1^2m_2^2}/(2M)$
denotes the final state's three-momentum in the rest frame of the
decaying particle. The experimental value of $R$ in the $J/\psi$ case
\cite{PDG98} has already been used in the phenomenological
determination of the basic mixing parameters in \cite{FeKrSt98}. Using
the phenomenological value of $\theta_8$ quoted in Table~\ref{Comptab},
we predict $R(\psi')=5.8$ and $R(\Upsilon)=6.5$. The prediction for
$R(\psi')$ agrees with the experimental value $2.9 ^{+5.4}_{-1.8}$,
which still has uncomfortably large errors however \cite{Bai:1998ny}.
For the radiative $\Upsilon$ decays only upper bounds exist at present.

\vskip1em\noindent {\it $\chi_{cJ}$ decays into two pseudoscalars:}

Because these are energetic decays the current quarks produced will be 
in an almost pure SU(3) singlet state. However, flavor symmetry 
violation can occur in the hadronization process. 
The ratio of $\chi_{cJ}$ decay widths into different
pairs of pseudoscalar mesons can be written ($J=0,2$)
\beq
\frac{\Gamma[\chi_{cJ}\to P_1 P_2]}{\Gamma[\chi_{cJ}\to P_3 P_4]}
&=&
\left(\frac{C_{12}}{C_{34}} \right)^2 \,
\left(\frac{k_{12}}{k_{34}}\right)^{2J+1}
\label{cij} \ .
\eeq
For the coefficients $C_{ij}$ two limiting cases can be considered.
If the mesons are formed at hadronic distances the influence of the
different decay constants will be a minor one and one expects to a
good accuracy
$ C_{\eta \eta}= C_{\eta' \eta'}=C_{\pi^0 \pi^0}$,
 $C_{\eta\eta'}=0$ and, from isospin symmetry,
 $C_{\pi^+\pi^-}=\sqrt 2 C_{\pi^0 \pi^0}$ (where we included the
statistical factor $\sqrt 2$).
If, however, the meson generation starts already at a time at
which the inter-quark distances are very small,
the decay amplitudes are obtained from the convolution of a hard scattering
process with the corresponding wave functions \cite{Bolz:1997ez}.
Assuming equal shapes of the wave functions, an assumption which
is not in conflict with present experimental information,
differences in the decay amplitudes are then solely due to the 
different decay constants and the mixing angle. One finds:
$C_{\eta\eta}= f_q^2 \, \cos^2\phi + f_s^2 \, \sin^2\phi
=1.41\,C_{\pi^0\pi^0}$,
$C_{\eta'\eta'}= f_q^2 \, \sin^2\phi + f_s^2 \, \cos^2\phi = 1.53 \,
C_{\pi^0\pi^0}$ and (with the statistical factor $\sqrt 2$ included)
$C_{\eta\eta'}= \sqrt 2  (f_q^2 - f_s^2) \, \sin \phi \, \cos\phi= -0.45 \,
C_{\pi^0\pi^0}$ and  $C_{\pi^+\pi^-}=\sqrt 2 C_{\pi^0 \pi^0}$.
Numerically we obtain
$\Gamma[\chi_{c0(2)}\to \eta  \eta]/
 \Gamma[\chi_{c0(2)}\to \pi^0 \pi^0]=1.9 ~(1.7)$,
$\Gamma[\chi_{c0(2)}\to \eta' \eta']/
 \Gamma[\chi_{c0(2)}\to \pi^0 \pi^0]=1.9 ~(1.3)$,
$\Gamma[\chi_{c0(2)}\to \eta  \eta']/
 \Gamma[\chi_{c0(2)}\to \pi^0 \pi^0]=0.2 ~(0.1)$ and
$\Gamma[\chi_{c0(2)}\to \pi^+ \pi^-]/
\Gamma_[\chi_{c0(2)}\to \pi^0 \pi^0]= 2  $.

From the differences between the two limiting cases it appears that
$\chi_{cJ}$ decays are less suited for testing
$\eta$-$\eta'$ mixing parameters, but, taking
the mixing parameters from other processes, they will provide
interesting information on meson formation in these reactions.
Experimentally,
only the ratio $\Gamma[\chi_{c0(2)}\to \eta  \eta]/
 \Gamma[\chi_{c0(2)}\to \pi^0 \pi^0]$ is known \cite{PDG98,BES-XX}:
$ 0.76^{+1.1}_{-0.5}~~(0.76^{+0.9}_{-0.6})$
(for the $\pi^0\pi^0$ branching ratio we combined the data with the
one for the $\pi^+\pi^-$ channel).
At present, the large experimental errors prevent any
definite conclusion.

Similar relations as given here (modified according to the correct charge
factors) should hold for two-photon annihilations into pairs of
pseudoscalar mesons.

\vskip1em\noindent {\it $g^*g^* \to \eta,\eta'$ transition form
factors:}

These form factors offer, in principle, a way to measure the angle
$\theta_1$. Allowing both the gluons to be virtual where at least
one of the virtualities $q_1^2$ and $q_2^2$ is supposed to be
very large, one may easily work out the leading-twist result for
these form factors \cite{BrodskyLepage}. In this approximation
one has ($i=q,s$)
\beq 
F_{\eta_i g^*}(q_1^2,q_2^2) &=& 
- C_i \, \alpha_s \, f_i \, \int dx \, 
  \frac{\phi_{i}(x)}{x \, q_1^2 + (1-x) \, q_2^2} 
\eeq
where $ \phi_{i}$ is the $\eta_i$ distribution amplitude, and $C_i$ a
numerical factor ($C_q=\sqrt2$, $C_s=1$). Combining both the form
factors into those for the physical mesons and assuming the equality
of the two distribution amplitudes (which at least holds in the formal
limit $q_i^2\to\infty$ since both the distribution amplitudes evolve
into the asymptotic one), one arrives at 
\beq 
\frac{F_{\eta\phantom{{}'} g^*}(q_1^2,q_2^2)}
     {F_{\eta' g^*}(q_1^2,q_2^2)}
&=& 
\frac{\sqrt 2 \, f_q \, \cos\phi - f_s \, \sin\phi} 
     {\sqrt2 \, f_q \, \sin\phi + f_s \, \cos\phi} 
= - \tan\theta_1 \ . 
\eeq 
Of course, at finite values of momentum transfer one may expect
corrections from differences between the two distribution amplitudes
and from transverse momentum. Such corrections can be worked out
following, for instance, Ref.~\cite{FeKrFF}.

For a measurement of these form factors one may consider the process
$pp\to {\rm jet + jet} + \eta(\eta')$ where the mesons are
supposed to be produced in the central rapidity region
\cite{Frere:1998ez}. According to Close \cite{Close:1997us} these
form factors may also be of relevance in $pp \to pp\eta(\eta')$,
provided the Pomeron couples to quarks $\propto \gamma^\mu$ and/or
$gg \to \eta(\eta')$ is the elementary process in the
Pomeron-Pomeron interaction. The explicit extraction of the form
factors from such measurements may, however, be very difficult.
Kilian and Nachtmann \cite{Kilian:1997ew} discuss
$\gamma$-Odderon-$\eta(\eta')$ form factors which appear in
diffractive $e$-$p$ scattering processes. We find that the ratio
of these form factors is given by $\cot\theta_8$ at large momentum
transfer.

\vskip1em\noindent {\it $Z\to\eta(\eta')\gamma$ decay:}

The treatment of these processes is rather academic since the expected
branching ratios are far below the present experimental bounds (The
$Z\to\pi\gamma$ decay has e.g.\ been discussed in
Ref.~\cite{Manohar:1990hu}). Nevertheless, they provide an additional
example of reactions which are sensitive to the angle $\theta_1$. The
ratio of the decay widths are given by  
\beq  
&& 
R(Z)= \frac{\Gamma[Z \to \eta'\gamma]}{\Gamma[Z \to \eta\gamma]} =
  \left|\frac{F_{\eta'\gamma Z}}{F_{\eta\gamma Z}}\right|^2 \,
  \left(\frac{k_{\eta'\gamma}}{k_{\eta\gamma}}\right)^3  
\eeq  
where
$F_{P\gamma Z}$ is the time-like form factor for $P\gamma$ transitions
mediated by the $Z$ boson. That form factor can be calculated along
the same lines as the $P\gamma\gamma^*$ transition form factor
\cite{FeKrSt98,FeKrFF}. Since the value of $M_Z^2$ is very large it
suffices to consider the asymptotic limit of the form factor only. In
terms of the octet and singlet decay constants, defined in
Eq.~(\ref{tas}), the result reads:  
\beq  
  F_{P\gamma Z}(M_Z^2)  &=&
  \frac{ 6 \, C_{8 \gamma Z} \, f_P^8 + 6 \, C_{1 \gamma Z} \, f_P^1}
       {M_Z^2}  
\eeq  
where $C_{8\gamma Z} = (1-4\sin^2\theta_W)/(6\sqrt6)$
and $C_{1\gamma Z} = (2-4\sin^2\theta_W)/(3\sqrt3)$. The weak coupling
of the flavor octet current is strongly suppressed by
$(1-4\sin^2\theta_W)$. Hence, $R(Z)\simeq\cot^2\theta_1$.

The same transition form factors $F_{P\gamma Z}$ appear in
$\eta(\eta') \to \gamma\mu^+\mu^-$ decays, but the momentum transfer
is very small. In analogy to the $P\to\gamma\gamma$ decays the
amplitudes in this case involve the inverse decay constants, and the
$\eta$ to $\eta'$ ratio is sensitive to the angle $\theta_8$.  To
measure these form factors in $\eta(\eta') \to\gamma\mu^+\mu^-$ decays
one has to extract the $\gamma$-$Z$ interference term from suitably
chosen asymmetries as discussed in detail in \cite{Bernabeu:1998hy}.

\vskip1em\noindent {\it Radiative transitions between light
 vector and pseudoscalar mesons:}

The relevant coupling constants are defined by \cite{Dumbrajs:1983jd}
\beq 
\langle P(p_P)| J_\mu^{\rm EM} | V(p_V,\lambda)\rangle |_{q^2=0}
&=&  
- g_{VP\gamma} \, \epsilon_{\mu\nu\rho\sigma} \, p_P^\nu \,
  p_V^\rho \varepsilon^{\sigma}{(\lambda)} \ .  
\eeq In
Ref.~\cite{BaFrTy95} these coupling constants are expressed in terms
of meson masses and decay constants by exploiting the chiral anomaly
prediction (at $q^2=0$) and vector meson dominance. However, the
difference between $\theta_8$ and $\theta_1$ has not been considered.
Translating the expressions for $g_{PV\gamma}$ correctly to the
quark-flavor scheme, following otherwise Ref.~\cite{BaFrTy95}, we
arrive at the formulas and values listed in Table~\ref{coupltab}.

The numerical result for the coupling constants depend on the actual
values of the vector mixing angle $\phi_V$, which is expected to
amount to only a few degrees (see e.g.\ \cite{Bramon97,BaFrTy95}).
The values in Table~\ref{coupltab} are calculated for $\phi_V=0$.  For
the vector meson decay constants we take \cite{Neubert:1997uc}
$f_\rho=210 $~MeV, $f_\omega=195 $~MeV,  $f_\phi=237 $~MeV.  The
predictions agree rather well with experiment. Indeed the relations
\beq 
\left| \frac{g_{\rho\eta'\gamma}}{g_{\rho\eta\gamma}}\right| =
\left| \frac{g_{\omega\eta'\gamma}}{g_{\omega\eta\gamma}}\right| =
\left| \frac{g_{\phi\eta\gamma}}{g_{\phi\eta'\gamma}}\right| 
&=& 
\tan\phi = 0.82
\eeq 
are well confirmed by experiment.  In Ref.~\cite{BaFrTy95}
results of similar quality could only be achieved by  using a value of
$\theta=\theta_8=\theta_1=-17^\circ$ which deviates from the values
obtained from other applications substantially.

\begin{table*}[hbt]
\begin{center}
\begin{tabular}{ll|lll}
$P$ & $V$ & $g_{VP\gamma}$ (in units $\frac{m_V}{f_V\pi^2}$) &
          $|g_{VP\gamma}|$(theo.) & $|g_{VP\gamma}|$(exp.) \\
\hline
$\eta$ & $\rho$ &
$\frac{3 \,\cos\phi}{4 \, f_q}$ & 1.52 & $1.85\pm 0.34$ 
\\
$\eta'$ & $\rho$ & 
$\frac{3 \, \sin\phi}{4 \, f_q}$ & 1.24 &$1.31\pm 0.12$ 
\\
$\eta$ & $\omega$ &
$ \frac{\cos\phi\cos\phi_V}{4 \, f_q}  
- \frac{2\, \sin\phi\sin\phi_V}{4 \, f_s}$ & 0.56 & $0.60\pm 0.15$
\\
$\eta'$ & $\omega$ & 
$ \frac{\sin\phi\cos\phi_V}{4 \, f_q}
+ \frac{2\, \cos\phi\sin\phi_V}{4 \, f_s}$ & 0.46 & $0.45\pm 0.06$ 
\\
$\eta$ & $\phi$ & 
$ \frac{\cos\phi\sin\phi_V}{4 \, f_q}
+ \frac{2\, \sin\phi\cos\phi_V}{4 \, f_s}$ & 0.78 & $0.70\pm 0.03$ 
\\
$\eta'$ & $\phi$ & 
$ \frac{\sin\phi\sin\phi_V}{4 \, f_q}
- \frac{2\, \cos\phi\cos\phi_V}{4 \, f_s}$ & 0.95 &$1.01 \pm 0.25$
\end{tabular}
\end{center}
\caption[]{Various coupling constants $g_{VP\gamma}$ from theory and
experiment \cite{PDG98}.  The numerical values are quoted in units of
GeV$^{-1}$.}
\label{coupltab}
\end{table*}

\vskip1em\noindent {\it $\eta$ and $\eta'$ admixtures to the pion:}

As is well-known (see e.g.\ \cite{Leutwyler96})  an accurate
prescription of the decays of $\eta(\eta')$ to three pions can only be
achieved by taking isospin violation into account. This effect is
usually  parametrized in terms of $\eta$ and $\eta'$ admixtures to
the pion,
\beq 
\pi^0 
&=& \phi_3 + \epsilon \, |\eta\rangle + \epsilon' \, |\eta'\rangle 
\eeq 
where $\phi_3$ denotes the pure isospin-1 state.
A straightforward generalization of our mixing scheme yields for the
strength of $\eta$ and $\eta'$ admixtures in the pion  
\beq 
&&
\epsilon  = \cos\phi \, \frac{m_{dd}^2 - m_{uu}^2}
                             {2 \, (M_\eta^2 - M_\pi^2)} 
\ , \qquad 
\epsilon'  = \sin\phi \, \frac{m_{dd}^2 - m_{uu}^2}
                             {2 \, (M_{\eta'}^2 - M_\pi^2)}
\label{epsepsp} \ , 
\eeq where the difference $m_{dd}^2 - m_{uu}^2$ can be estimated from
$2 (M_{K^0}^2 - M_{K^\pm}^2 + M_{\pi^\pm}^2 - M_{\pi^0}^2)$ to amount
to $0.0104$~GeV$^2$.  A possible difference in $u-$ and $d-$ quark
decay constants is ignored in the derivation of (\ref{epsepsp}). The
expressions for $\epsilon$ and $\epsilon'$ look rather simple in the
quark flavor scheme and are intimately connected to physical
quantities. Inserting our phenomenological number for the mixing angle
$\phi$ we obtain $\epsilon=0.014$ and $\epsilon'=0.0037$.  By
exploiting the properties of the mass matrix \cite{FeKrSt98} the ratio
$\epsilon/\epsilon'$ following from (\ref{epsepsp}) can also be
expressed in terms of $\theta_8$ and $\theta$
\beq
\frac{\epsilon'}{\epsilon} 
&=& 
- \tan\theta_8 \, 
\left(\frac{ \cos\theta + \sqrt2 \, \sin\theta} 
                             {\sqrt2 \, \cos\theta - \sin\theta}
\right)^2 \ .
\label{epspeps}
\eeq The numerical value, following from our phenomenological set of
parameters, is $0.26$.  In contrast, the conventional approach (see
e.g.\ \cite{Leutwyler96}), using $\theta=\theta_8\simeq -20^\circ$
gives the much smaller value $0.17$.

The values of the parameters $\epsilon$ and $\epsilon'$ have recently
been shown to be of importance for the investigation of $CP$-violation
in $B\to \pi\pi$ decays since it breaks the isospin triangle relation
for the  amplitudes of the three processes $B^+\to\pi^+\pi^0$,
$B^0\to\pi^0\pi^0$ and $B^0\to\pi^+\pi^-$ \cite{Gardner:1998gz}.  The
value of $\epsilon'$ used in Ref.\ [29] ($\epsilon'=0.0077$) is
substantially larger than our value.

\vskip1em\noindent{\bf Summary}

We discussed the mixing properties of the $\eta$ and $\eta'$ meson
state vectors and of their decay constants and showed that there is,
at most, only one basis where the mixing of the decay constants can
follow the pattern of state mixing.  Chiral perturbation theory as
well as phenomenological analyses favor this proposition for the
quark-flavor basis.  In general, e.g. in the familiar octet-singlet
basis, one needs two angles in order to parametrize the decay
constants.  However, when using our quark-flavor mixing scheme, these
new angles are fixed by the basic parameters $\phi$, $f_q$, $f_s$,
leading to a number of important consequences for many
reactions.  The
results are quite different from conventional mixing schemes in which
the subtleties discussed here are not considered and where, as a
consequence of that, the mixing parameters often show a strong process
dependence.
  
The improved knowledge of the mixing parameters is also of
importance for the analysis of $B$~decays, like $B\to K\eta'$ (see
e.g.\ \cite{Ali97b}) or $B\to\pi\pi$ \cite{Gardner:1998gz}.  Further
interesting applications of our approach refer to $\eta$ and $\eta'$
production processes  in high energy hadron collisions where exotic
form factors such as the  $g^*g^*\eta(\eta')$ form factors play an
important role.

\section*{Acknowledgements}

T.F.\ was supported by {\it Deutsche Forschungsgemeinschaft.}
P.K. thanks the {\it Special Research Centre for the Subatomic
Structure of Matter} at the University of Adelaide for support and
the hospitality extended to him.


\end{document}